\documentclass[twocolumn,usenatbib]{mnras}
\usepackage{url}
\usepackage{rotating}
\usepackage[normalem]{ulem} % for strikeout
\usepackage{graphicx}
\usepackage{pdflscape}		% For landscape table
\usepackage{amssymb} 		% For math symbols
\usepackage{tablefootnote} 	% For table footnote

\newcommand{\suz}{\emph{Suzaku}} 
\newcommand{\nus}{\emph{NuSTAR}} 
\newcommand{\swift}{\emph{Swift}} 
\newcommand{\obs}[1]{\textit{Obs~#1}}
\newcommand{\sw}[1]{\texttt{#1}}

\newcommand{\cep}{Cep~X-4}
\newcommand{\msun}{\ensuremath{M_\odot}}

\defcitealias{felix}{FF15}
\newcommand{\felix}{\citetalias{felix}}
\newcommand{\pfelix}{\citepalias{felix}}

\begin{document}

\title[Explaining the asymmetric line profile in Cepheus~X-4 ]{Explaining the asymmetric line profile in Cepheus~X-4 with spectral variation across pulse phase}

\author[Bhargava et al.]{Yash Bhargava$^1$\thanks{$\!$ {\vspace{-5pt}\small Email: yash@iucaa.in}}, Varun Bhalerao$^2$, Ralf Ballhausen$^3$, Felix F\"{u}rst$^4$,\newauthor Katja Pottschmidt$^{5,6}$, John A. Tomsick$^7$ and Joern Wilms$^3$ \\ 
\noindent$\!\!$ $^1$ Inter-University Centre for Astronomy and Astrophysics, Pune, India,\\
\noindent$\!\!$ $^2$ Indian Institute of Technology Bombay, Mumbai, India,\\
\noindent$\!\!$ $^3$ Dr. Karl Remeis-Sternwarte and Erlangen Centre for Astroparticle Physics, Sternwartstrasse 7, D-96049 Bamberg, Germany,\\
\noindent$\!\!$ $^4$ European Space Astronomy Centre (ESAC), Science Operations Departement, 28692 Villanueva de la Ca\~nada, Madrid, Spain,\\
\noindent$\!\!$ $^5$ CRESST, Department of Physics, and Center for Space Science and Technology, University of Maryland Baltimore County, \\ $\!\!$ 1000 Hilltop Circle, Baltimore, MD 21250, USA,\\
\noindent$\!\!$ $^6$ NASA Goddard Space Flight Center, Astrophysics Science Division, Code 661, Greenbelt, MD 20771, USA,\\
\noindent$\!\!$ $^7$ Space Sciences Laboratory, 7 Gauss Way, University of California, Berkeley, CA 94720-7450, USA}

\date{Accepted 2018 October 16. Received 2018 October 16; in original form 2018 April 05}
\maketitle

\begin{abstract}
The high mass X-ray binary \cep, during its 2014 outburst, showed evidence for an asymmetric cyclotron line in its hard X-ray spectrum. The 2014 spectrum provides one of the clearest cases of an asymmetric line profile among all studied sources with Cyclotron Resonance Scattering Features (CRSF). We present a phase-resolved analysis of \nus\ and \suz\ data taken at the peak and during the decline phases of this outburst. We find that the pulse-phased resolved spectra are well-fit by a single, symmetric cyclotron feature. The fit parameters vary strongly with pulse phase: most notably the central energy and depth of the cyclotron feature, the slope of the power-law component, and the absorbing column density. We synthesise a ``phase averaged'' spectrum using the best-fit parameters for these individual pulse phases, and find that this combined model spectrum has a similar asymmetry in the cyclotron features as discovered in phase-averaged data. We conclude that the pulse phase resolved analysis with simple symmetric line profiles when combined can explain the asymmetry detected in the phase-averaged data.
\end{abstract}


\section{Introduction}

High Mass X-ray Binaries (HMXBs) are systems containing a compact object --- either a neutron star (NS) or a black hole --- gravitationally bound to a massive, early type star (M $>\ 5$~\msun). In accreting NS HMXBs, the high magnetic field ($\gtrsim 10^{12}$~G) of the NS channels the accreted matter onto the magnetic poles \citep[and references therein]{Mukherjee2013MNRAS.430.1976M}. The kinetic energy of the accreted matter is converted to X-rays near the surface of the NS. X-ray spectroscopy thus allows us to probe regions close to the neutron star. Of particular interest is the effect of the magnetic field on the X-ray spectrum. Resonant scattering of X-ray photons off electrons occupying and transitioning between quantised Landau levels creates Cyclotron Resonant Scattering Features (CRSFs) in the X-ray spectra. The central energy of these features relates directly to the local magnetic field as $E_{\rm cyc} =  11.6~B_{12} (1+z)^{-1}$~keV, where $B_{12}$ is magnetic field in units of $10^{12}$~G and $z$ is the gravitational redshift.

CRSFs have been observed in several NS HMXBs \citep[see for example][]{Caballero2012MmSAI..83..230C, Mukherjee2013MNRAS.430.1976M, Jaisawal2017symm.conf..153J}. These cyclotron features and their harmonics, if any, are usually modelled with a Gaussian optical depth line \sw{gabs} or with the psuedo-Lorentzian profile \sw{cyclabs} phenomenological model \citep{Mihara1990Natur.346..250M, Makashima1990PASJ...42..295M}. The asymmetric profiles give hints towards the viewing angle onto the magnetic field and the field configuration \citep[see][]{Schwarm2017A&A...597A...3S, Schwarm2017A&A...601A..99S}. While theoretical work has been done on the expected complex shapes of these features \citep[and references therein]{Mukherjee2012MNRAS.420..720M, Mukherjee2013MNRAS.430.1976M, Schwarm2017A&A...597A...3S, Schwarm2017A&A...601A..99S}, further studies have been limited by the energy resolution and sensitivity of X-ray telescopes. Due to excellent spectral resolution of \nus\ \citep[0.4~keV at 10~keV,][]{Harrison2013ApJ...770..103H}, the shapes of CRSFs can be probed by \nus.

The HMXB Cepheus X-4 (hereafter \cep) was discovered in outburst in 1972 by the \emph{OSO-7} satellite~\citep{Ulmer1973ApJ...184L.117U}. Further studies of a 1988 observation led to the discovery of 66.3~s pulsations~\citep{Koyama1991ApJ...366L..19K} and identification of the optical B[e] companion~\citep{Roche1997IAUC.6698....2R,bm98}. \citet{Wilson1999ApJ...511..367W} used the outbursts seen by \emph{BATSE} and \emph{RXTE} in 1993 and 1997 to constrain the orbital period to be between 23 and 147.3 days. \citet{McBride2007A&A...470.1065M} used long term \emph{RXTE}-ASM lightcurve to constrain the orbital period of \cep\ to $20.85\pm0.05$~days. \cite{mmk+91} reported the first detection of the cyclotron feature at $30.5\pm0.4$~keV in \cep, which was confirmed by \citet{McBride2007A&A...470.1065M} with \emph{RXTE} observations of the 2002 outburst. 
Detailed \nus\ observations by \citet{felix} (hereafter \felix) during the 2014 outburst of \cep\ provided clear evidence of an asymmetric CRSF profile in the phase-averaged spectrum, which we explore further in this work. \cite{Schwarm2017A&A...601A..99S} successfully described the asymmetric line profile observed with \nus\ at the peak of this outburst with their physical model based on Monte Carlo modelling of CRSFs. Analyses by \cite{Jaisawal2015MNRAS.453L..21J} of a lower flux \suz\ observation of this outburst and by \cite{Vybornov2017A&A...601A.126V} of the \nus\ observations led to the discovery of a harmonic CRSF with an energy ratio to the fundamental line of 1:1.7 and 1.8, respectively.

As the NS rotates, X-rays from different regions near its surface become visible from Earth.  Variations of the  cyclotron features with pulse phase can be used to probe variations of the magnetic fields along different lines of sight. Variations of the cyclotron line parameters and of the continuum parameters over the pulse phase have been observed in many sources, showing the importance of phase-resolved analysis \citep[see, e.g.,][]{Heindl2004AIPC..714..323H, Maitra2017JApA...38...50M}. \citet{Jaisawal2015MNRAS.453L..21J} performed pulse phase resolved analysis for \cep, and have reported significant parameter variation.

In this work, we undertake detailed pulse phase-resolved analysis of \cep\ by combining quasi-simultaneous \nus\ and \suz\ data from the 2014 outburst. The organisation of this paper is as follows: we present the observations and data reduction procedures in \S\ref{sec:obs_data_red}. We extend the phase-averaged analysis of \felix\ to include the \suz\ data, and compare results in  \S\ref{subsec:phase_av}. In \S\ref{subsec:phaseres} we divide the data into 9 phase bins over the $\sim 66.3$~s period, and explore variations of spectral parameters across these bins. In particular, we examine evidence for any line asymmetry in these individual phase bins. Finally, in \S\ref{subsec:add_model} we test if the asymmetric line profile we previously reported \felix\ can be entirely explained by averaging the spectra from various phases. We conclude with a discussion in  \S\ref{sec:disc}.

\section{Observations and Data reduction} \label{sec:obs_data_red}

\begin{figure}\centering
\includegraphics[width=\columnwidth]{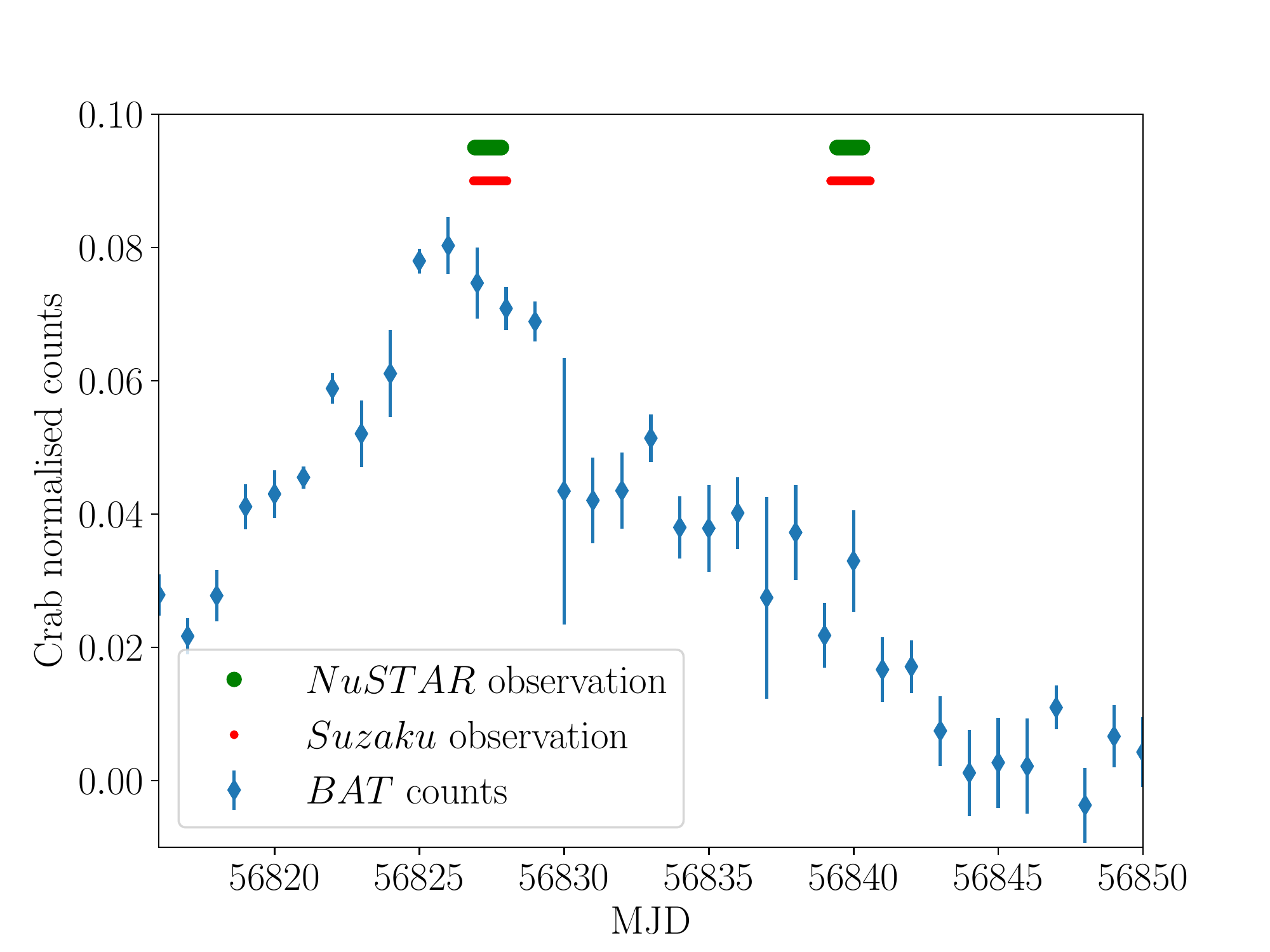}
\caption{\emph{Swift}-BAT lightcurve of the 2014 outburst of \cep\ (blue circles). The horizontal green and red lines near the top indicate periods of \nus\ and \suz\ observations respectively. In this work, we refer to the pair of observations near the peak of the outburst as \obs1, and to observations during the decline of the outburst ($\sim$MJD~56840) as \obs2.}

\label{fig:bat_lc}
\end{figure}

Motivated by the detection of an outburst in \cep\ \citep{Nakajima2014ATel.6212....1N},
we triggered target-of-opportunity observations with \nus\ and \suz\ in June--July 2014 (\autoref{tab:obs_list}). We obtained two sets of observations: near the peak of the outburst (hereafter \obs1) and during the decline (hereafter \obs2). The times of these quasi-simultaneous observations are shown superposed on the \emph{Swift}-BAT lightcurve in \autoref{fig:bat_lc}. These observations were also accompanied by short 1--1.6~ks snapshots with \swift~XRT~\pfelix. However, the very short exposures of these observations are not useful for pulse phase resolved analysis and hence are excluded from the current work.

\begin{table}
\caption{Summary of \nus\ and \suz\ observations of \cep\ used in this work}
\label{tab:obs_list}
\scriptsize
\begin{tabular}{l|c|c}
\hline
 & \obs1 & \obs2 \\
 \hline
Observation ID & & \\
\nus & 80002016002 & 80002016004\\
\suz & 409037010 & 909001010\\
\hline
MJD range & & \\
\nus & 56826.92 -- 56827.84 & 56839.43 -- 56840.31\\
\suz & 56826.87 -- 56828.03 & 56839.21 -- 56840.58 \\ \hline
Exp time (ks) & & \\
\nus & 40.4 & 41.1\\
\suz\ XIS3 & 30.9 & 60.4\\ 
\suz\ HXD/PIN & 50.4 & 75.7\\\hline

\end{tabular}
\end{table}

\subsection{\nus\ data reduction}

\nus\ data were reduced using the standard FTOOL \sw{nupipeline} software version 1.7.1 (HEAsoft version 6.22) and \nus\ CALDB version 20171002. Total exposure time after screening the data for \obs1\ was $\sim$40~ks and for \obs2\ was $\sim$41~ks. Source counts were extracted from a circular region of 120\arcsec\ radius centred on the source coordinates, while background counts were extracted from a circular region of 90\arcsec\ radius farthest away from the source. We applied barycentric correction with FTOOL \sw{barycorr} and extracted pulse phase resolved spectra using FTOOL \sw{xselect}. %

\subsection{\suz\ data reduction}

\suz\ data were reprocessed and screened applying standard criteria using \sw{aepipeline} (HEAsoft version 6.16). We used CALDB version 20150105 for XIS and version 20110913 for HXD. We performed attitude correction using FTOOL \sw{aeattcor2}. The data were corrected to the solar system barycenter. We extracted the XIS lightcurves and spectra using the FTOOL \sw{xselect} and HXD/PIN lightcurves and spectra using the \suz\ specific FTOOLS \sw{hxdpinxbpi} and \sw{hxdpinxblc} respectively.

Only XIS3 data could be used for the data analysis as XIS1 had a mode switch during \obs1 and XIS0 had incomplete and strongly misaligned image with low count rates, while XIS0 and XIS1 were terminated during \obs2. XIS3 was operated in 1/4 window mode during \obs1. 
Source photons were extracted from an annulus with an outer radius 100\arcsec\ for both observations. To avoid pileup effects of over 4\%, we excluded the central 45\arcsec\ and 10\arcsec\  regions in \obs1\ and \obs2\ respectively. For both of the observations we combined the data from $3\times3$ and $5\times5$ editing modes. The background counts were extracted from a circular region of radius 100\arcsec\ in the outer regions of the chip and sufficiently far away from the source.

\section{Spectral analysis}
We first undertake phase-averaged analysis for both \obs1\ and \obs2\ in \S\ref{subsec:phase_av}, to compare our \nus+\suz\ results with our older \nus+\swift\ results~\pfelix. Next, in \S\ref{subsec:phaseres}, we divide the data into nine phase bins and examine the variations of model parameters with pulse phase. Finally, we synthesise phase-averaged spectra from phase-resolved model spectra, and examine them for the asymmetric line profile in \S\ref{subsec:add_model}.

\subsection{Phase averaged analysis} \label{subsec:phase_av}

We adopt the continuum model used by \felix\  and \citet{McBride2007A&A...470.1065M}: an absorbed soft black-body and a harder power law~(\sw{PL}) spectrum with a Fermi-Dirac cut-off~(\sw{FDCUT}) \citep{Tanaka1986LNP...255..198T} of the form:
\begin{equation}
F(E) \propto E^{-\Gamma}\left(\frac{1}{1+e^{-(E-E_{\rm{cut}})/E_{\rm{fold}}}}\right)
\end{equation}
Absorption by neutral hydrogen was included using the \sw{XSPEC} \sw{phabs} model with the abundances from \citet{Wilms2000ApJ...542..914W} and using the cross sections from \citet{Vern1996ApJ...465..487V}.
 To improve the fit an absorption line was added at 30~keV. This line, detected by various authors \citep[see, e.g.,][]{McBride2007A&A...470.1065M}, is expected to be a CRSF. In concordance with \felix, we find evidence of a secondary absorption feature at 18~keV. 
The strength of this feature is lower than the 30~keV line, leading to our interpretation of this being an asymmetric feature of the primary 30~keV cyclotron line. We do not find any evidence for a second harmonic at higher energies.
The quality of fits improves further when we include the Fe~K$\alpha$ line as a Gaussian feature at $\sim6.4$~keV. With this final model, \texttt{phabs*gabs*gabs*(bbodyrad+gaussian+fdcut*powerlaw)} (Model 1), we get satisfactory fits with $\chi^2 = 4708$ with 4199 degrees of freedom in \obs1, and $\chi^2 = 4073$ with 3794 degrees of freedom in \obs2. Our parameter values from this \nus\ and \suz\ data set confirm the findings of \felix\ from \nus\ and \emph{Swift}-XRT data (\autoref{tab:ph_av_comp}).

\citet{Jaisawal2015MNRAS.453L..21J} and \citet{Vybornov2017A&A...601A.126V} reported the detection of a harmonic to the CRSF. The ratio of the harmonic energy to CRSF energy is found to be $\sim1.8$ which is lower than the expected value of 2. The data in the energy range above 50~keV are dominated by the background for both \nus\ and \suz. Thus the parameters of the harmonic line can be expected to depend on the uncertainty level of the background. The cyclotron line and the harmonic are well localised and excluding the harmonic does not affect the continuum \citep{Vybornov2017A&A...601A.126V}. Since the focus of the current work is to study the fundamental CRSF, we restrict the studied energy range to $<50$~keV for \nus\ and $<45$~keV for \suz\ XIS.

\subsection{Pulse phase resolved analysis}\label{subsec:phaseres}
We used \sw{efsearch} to measure a pulse period of 66.335$^{+0.003}_{-0.004}$~s and 66.333$^{+0.007}_{-0.011}$~s in \obs1\ and \obs2\ respectively. We then selected a sharp feature in intensity, visible in both observations, as phase 0. This corresponds to barycentric MJD~56826.000023 for \obs1\ and MJD~56839.000215 for \obs2. 
The energy resolved pulse profiles of the source are shown in \autoref{fig:en_res_pulse}. For both observations, the pulse profile varies with energy. Several features get smoothed over at higher energies. In particular, the sharp feature used to define phase 0 is not seen above 10~keV. We also note that the pulse fraction seems to slightly increase with energy up to the 20--40~keV band (seen from the variation in the root-mean-squared values of the pulsed profile), beyond which the data are very noisy and the pulse cannot be detected significantly. To investigate this energy dependence quantitatively, we undertook pulse phase-resolved spectroscopy.

\begin{figure}
\centering
\includegraphics[width=1\columnwidth]{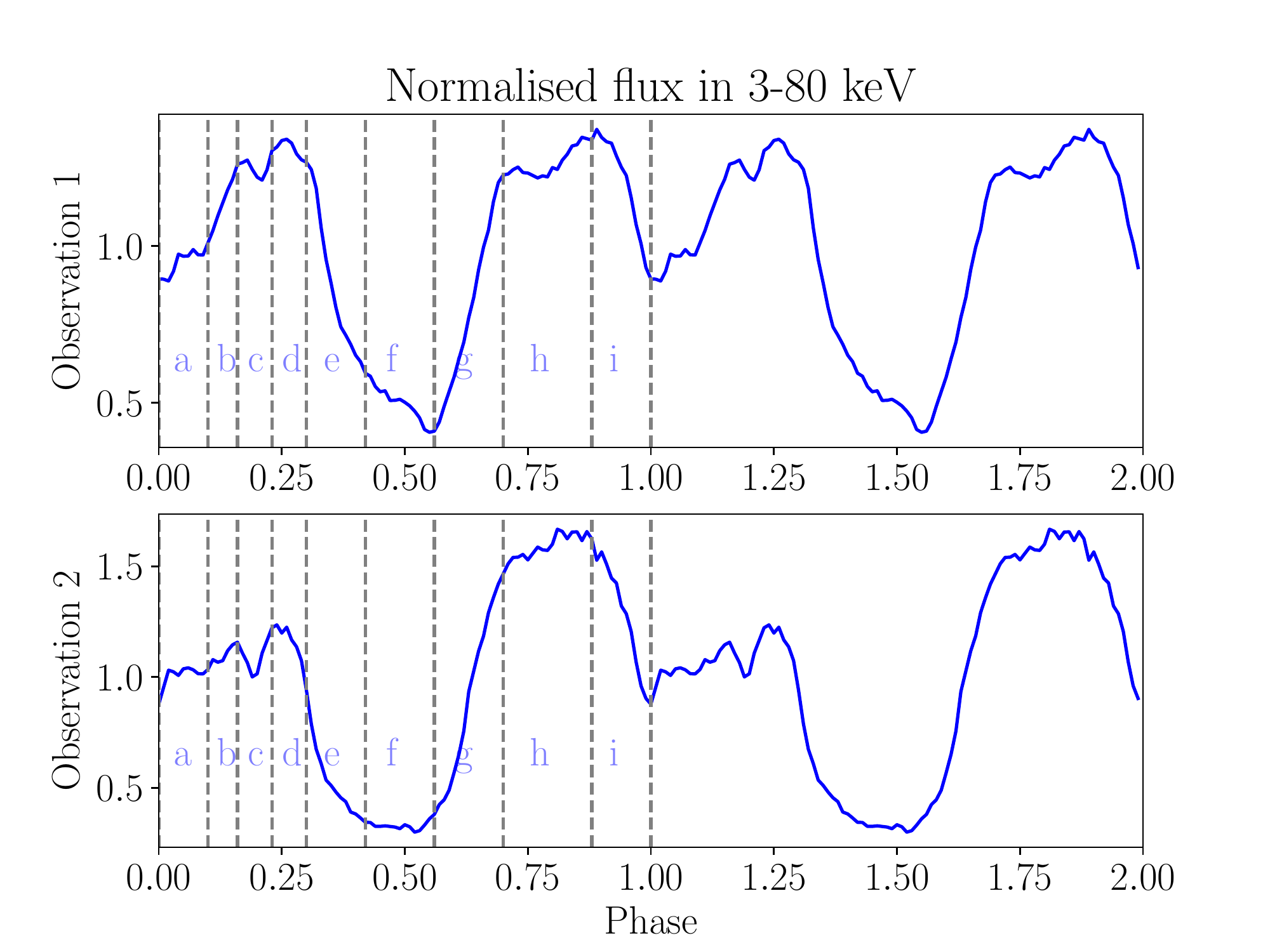}

\caption{Variation of 3--80~keV flux with pulse phase, as observed by \nus. Two cycles are shown for clarity. The top panel shows the the lightcurve for \obs1, folded on the pulse period of 66.335~s and divided into 100 uniformly spaced bins for plotting. Similarly, the bottom panel shows the lightcurve for \obs2\ folded on the corresponding period of 66.333~s. Vertical dashed lines shown in the first cycle, indicate the boundaries for the phase-resolved spectroscopy (\autoref{tab:all_phases}).
}

\label{fig:ph_bound}
\end{figure}

\begin{figure*}
\includegraphics[width=1\columnwidth]{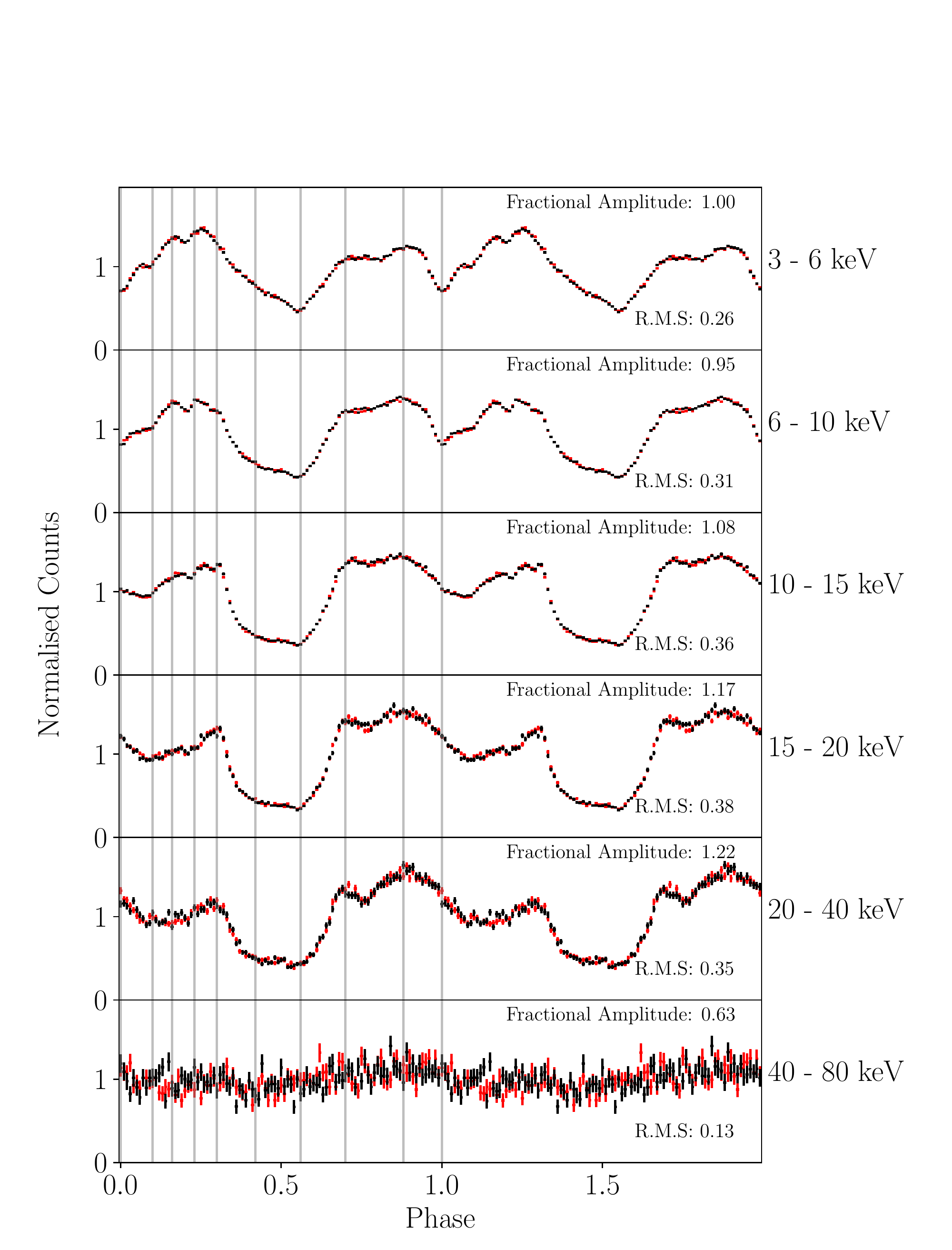}\includegraphics[width=1\columnwidth]{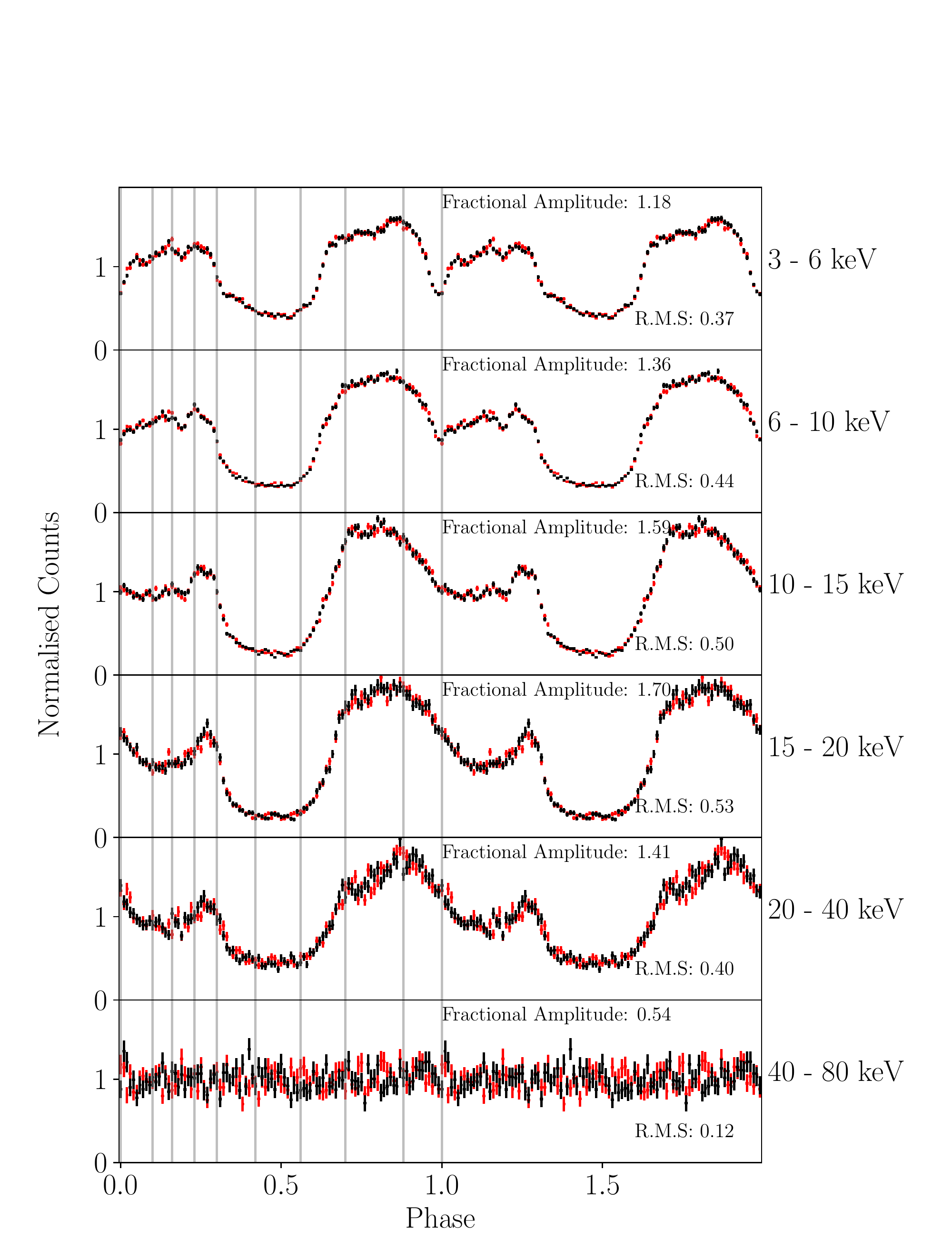}
\caption{Energy resolved pulse profiles of \obs1\ (Left panel) and \obs2\ (Right panel). The points in red correspond to \nus\ FPMA counts, while the points in black correspond to \nus\ FPMB counts, both normalised to the mean count rate in the respective energy bin. The y-axis for each box spans ranges from 0--2 in normalised units. Inset numbers show the root-mean-squared variability and the fractional amplitude across phase for each energy band. Vertical grey lines correspond to the phase boundaries used in the analysis in \S~\ref{subsec:phaseres}. The labelling of the phases is shown in \autoref{fig:ph_bound}, with boundaries defined in \autoref{tab:all_phases}.}

\label{fig:en_res_pulse}
\end{figure*}

We divided the pulse profile into 9 bins (a--i), as shown in \autoref{fig:ph_bound}. The phases of the boundaries are given in \autoref{tab:all_phases}. The phase boundaries were chosen at distinct transition points which include sharp changes in the local slope of the profile. Although \obs1\ and \obs2\ were analysed separately, the same phase boundaries were used for both. As background is not expected to vary significantly over pulse phase, for fitting each observation we used the respective phase-averaged background spectra. The spectra from all the instruments were rebinned to a minimum of 20 counts per bin for fitting purposes. 

\begin{figure*}
\includegraphics[width=0.48\textwidth]{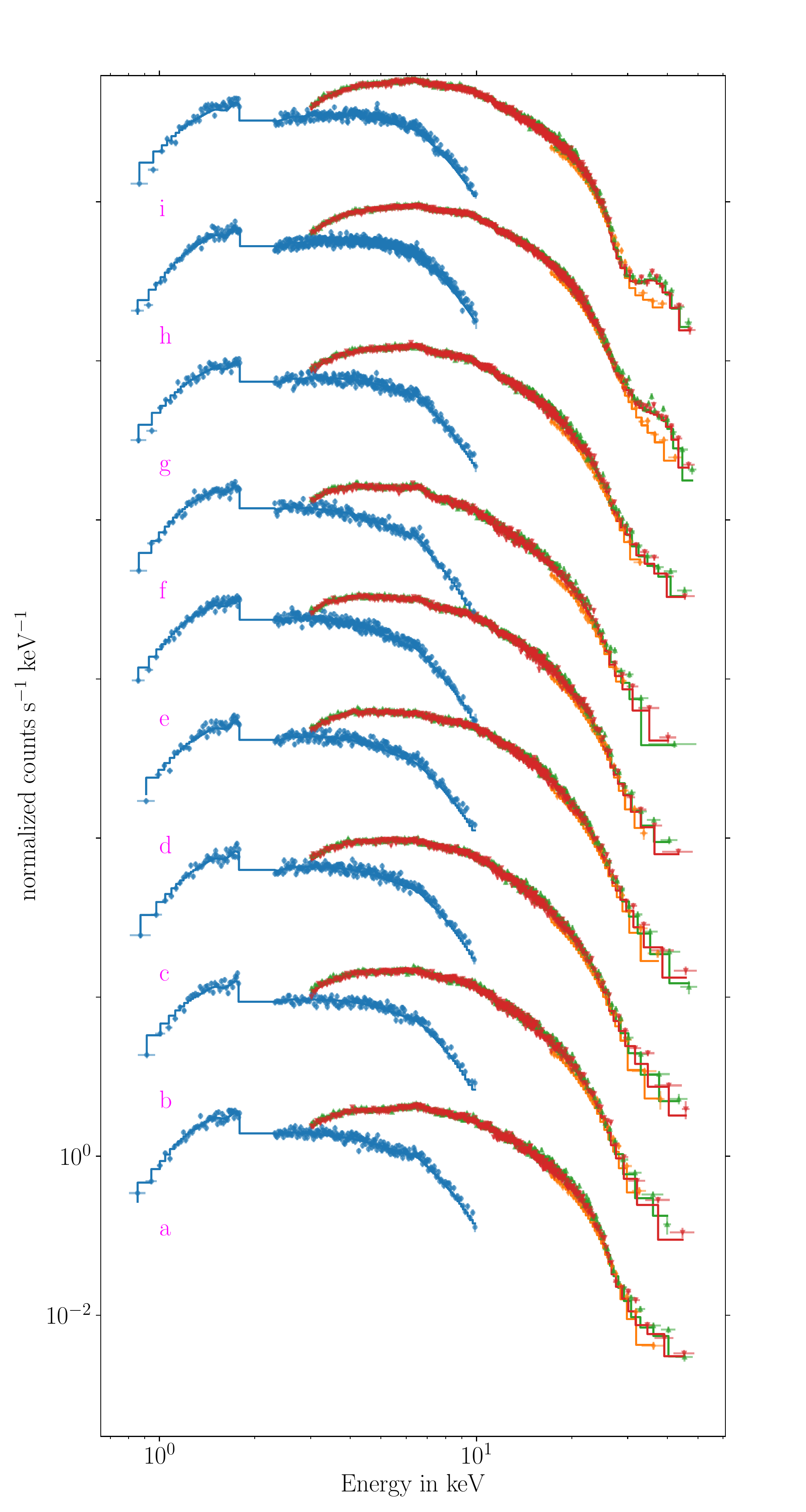}\includegraphics[width=0.48\textwidth]{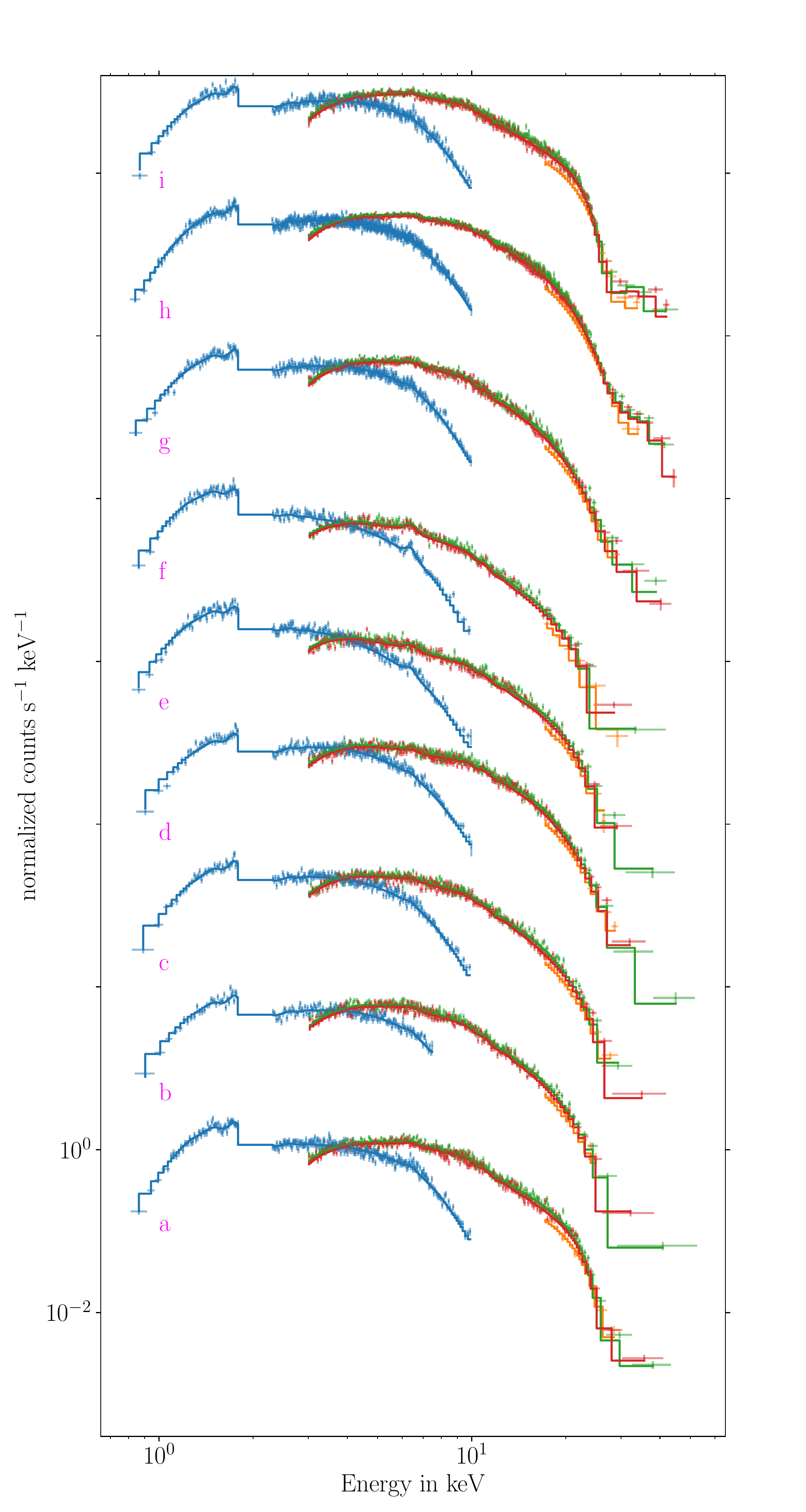} 
\caption{Spectrum for all the phases of \obs1\ (Left panel) and \obs2\ (Right  panel). Spectrum for phase a has been kept at the true value and the rest of the spectra are plotted with a constant offset (multiplied by 40). The blue circle, orange diamond, green triangle and red inverted triangle markers correspond to \suz--XIS, \suz--HXD, \nus--FPMA and \nus--FPMB spectra respectively. For plotting purposes only, the spectra were rebinned to have a significance of $10\sigma$ per bin, to clearly show the shape of the spectrum.  }
\label{fig:spec_all}
\end{figure*}

The spectra for individual phases have been plotted in \autoref{fig:spec_all}, (See left panel for \obs1 and right for \obs2). The consecutive spectra are scaled by a constant factor (40) to demonstrate the shape of the spectra. The spectrum for phase `a' (bottom spectrum in each panel) is unscaled. We fit each of these spectra with the spectral model \texttt{phabs*gabs*(bbodyrad+gaussian+fdcut*powerlaw)} (Model 2). The cross normalisations for the instruments were frozen to the phase averaged values. The quality of data in individual phases was not good enough to constrain the iron line parameters, hence these were kept frozen to the values obtained in phase-averaged spectroscopy (\S\ref{subsec:phase_av}). The freezing of iron line parameters led to increase of $\chi^2$ by up to $\sim$10 for all the phases, but we consider it acceptable as the ``best-fit'' line parameters obtained for individual phases are highly unphysical.

\begin{figure*}
\includegraphics[width=0.48\textwidth]{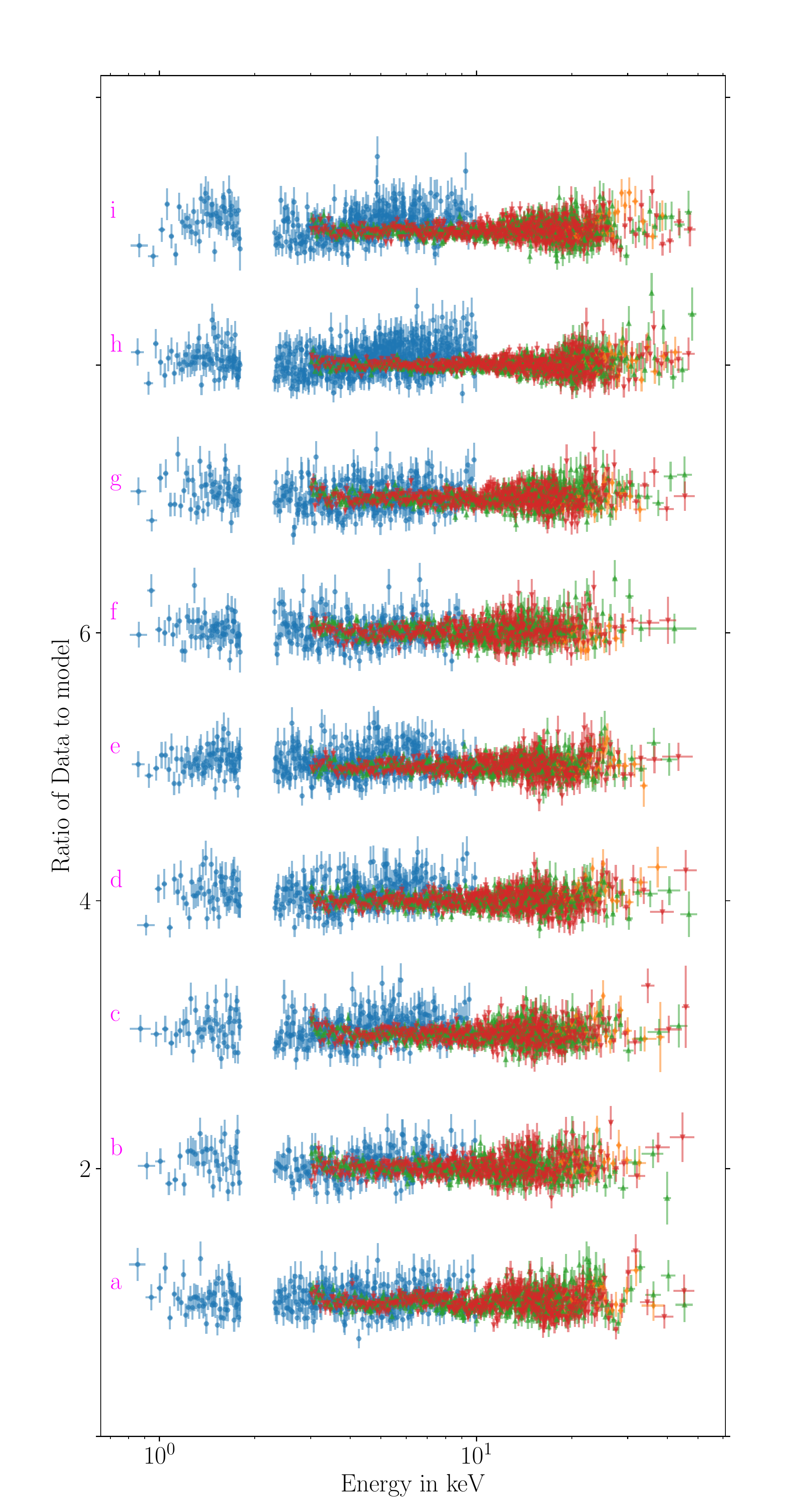}\includegraphics[width=0.48\textwidth]{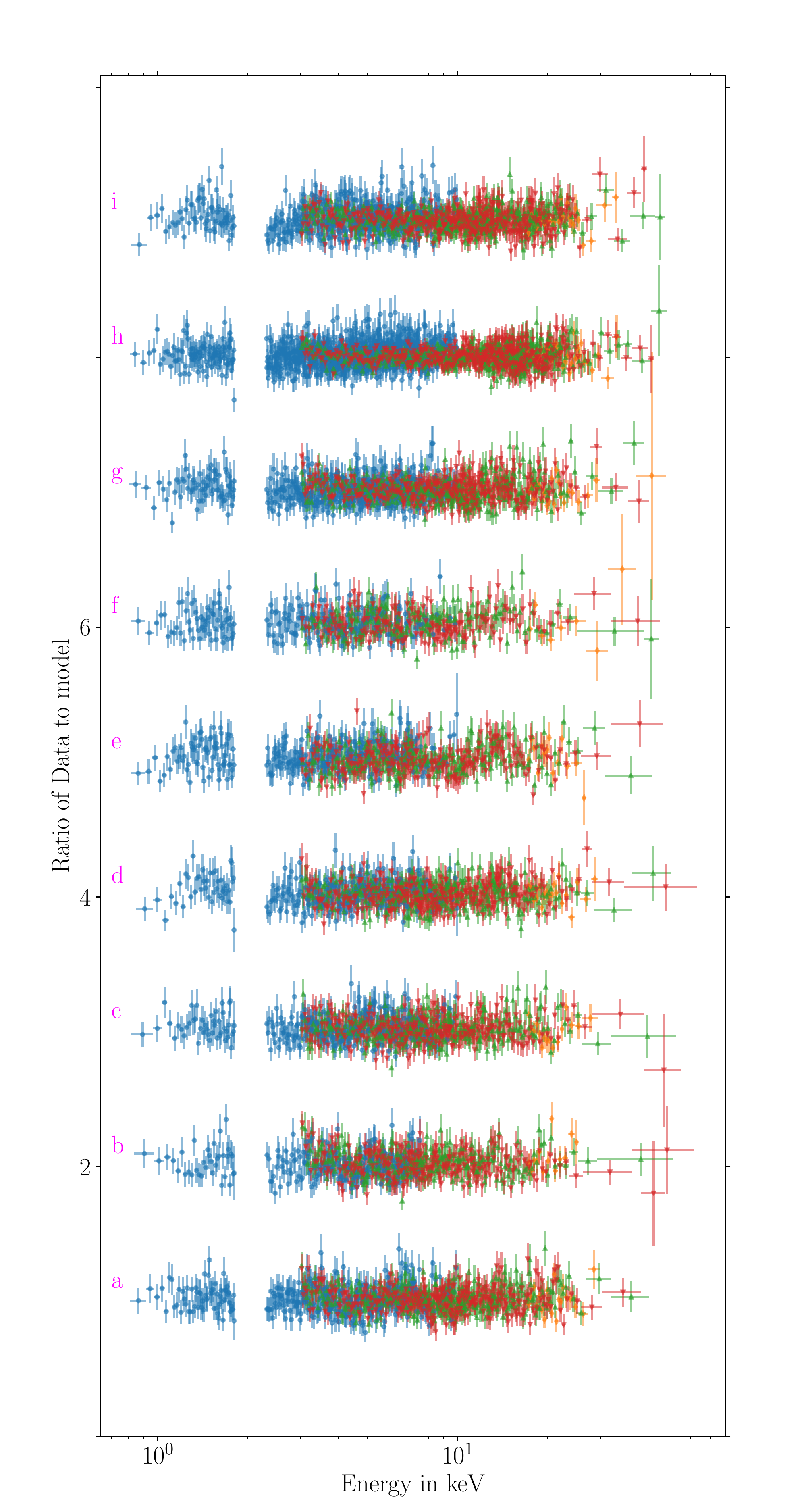} 
\caption{Residuals in form of ratio of data to model for all the phases of \obs1\ (Left panel) and \obs2\ (Right  panel). Residuals for phase a has been kept at the true value and the rest of the spectra are plotted with a constant offset (1). The colour scheme and the rebinning is same as \autoref{fig:spec_all}. The 18 keV feature seems non existent in the residuals but is clearly seen in the phase averaged case \pfelix.}
\label{fig:rat_all}
\end{figure*}

%We found no clear evidence of the asymmetric line profile in the spectra for individual phases. 
As attempted in \S\ref{subsec:phase_av} (and by \felix), another gaussian absorption line was added at 18~keV in the model for the phase bins, but it improved the $\chi^2$ by only small amounts ($<10$) for three additional parameters. This is also visually evident in the \autoref{fig:rat_all}, where the ratio of the data to model is shown for each of the phase bin. For clarity, the consecutive ratio plots are offset by a constant amount (1) and the ratio plot for phase `a' is kept at the actual value. An unmodelled absorption feature in the residuals, if present, would show up as a dip around the peak of the absorption feature, which is not seen in \autoref{fig:rat_all}. The significance of the additional gaussian absorption line was tested using Bayesian Information Criterion \citep[BIC, ][]{schwarz1978} since the \sw{ftest} included in \sw{XSPEC} cannot be used for nested multiplicative models \citep{Orlandini2012ApJ...748...86O}. A model with lower BIC is preferred as a better approximation of the true model. The evidence against a model with higher BIC can be estimated by the difference in BIC for the two models \citep{Kass1995BayesFactor}. BIC was computed  for the phases with highest reduction in $\chi^2$ from the single line model (Model~2) to the double-lined model (Model~1) (\obs1: Phase a,d; \obs2: Phase e,h). In all cases BIC is lower for Model 2 (single line). $\Delta$BIC ($\sim10$ for the tested phases) rules out the model with higher BIC strongly. Hence  we use Model 2 to describe the data from now on.

\begin{figure*}
\centering

\includegraphics[height=0.9\textheight]{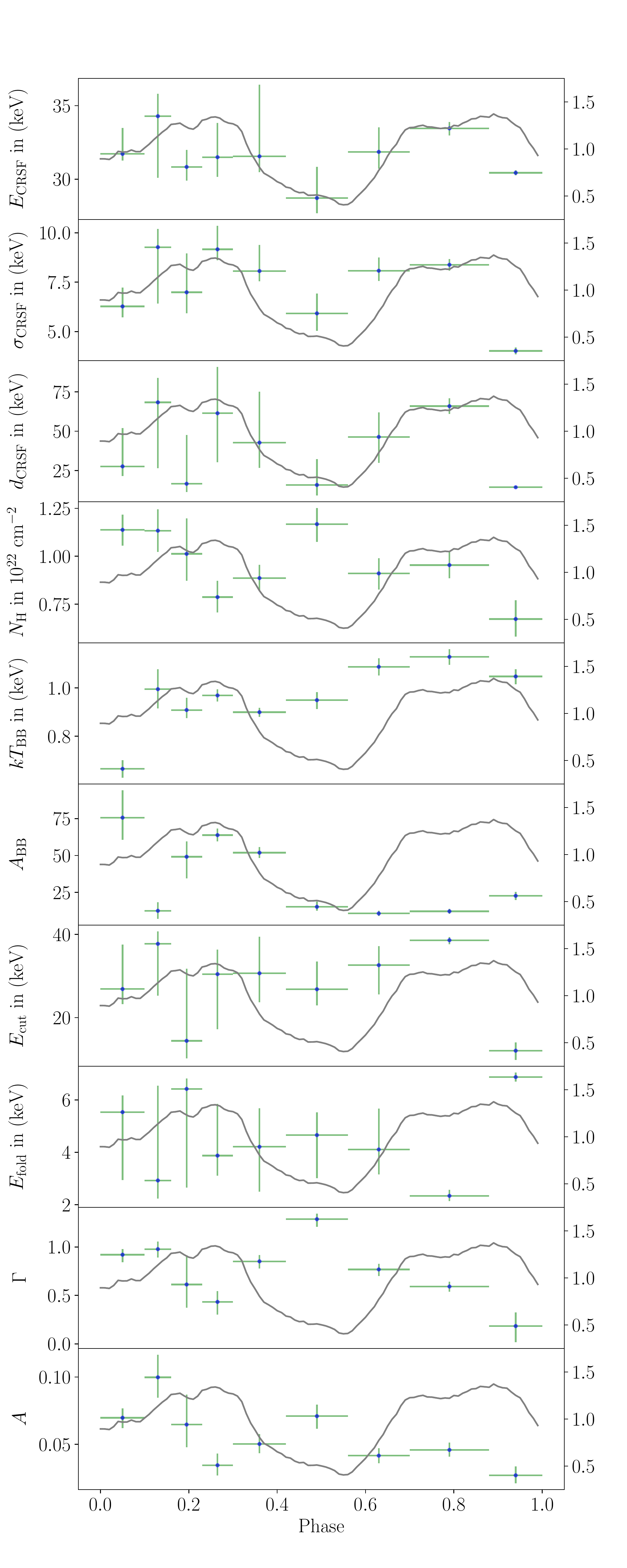}\includegraphics[height=0.9\textheight]{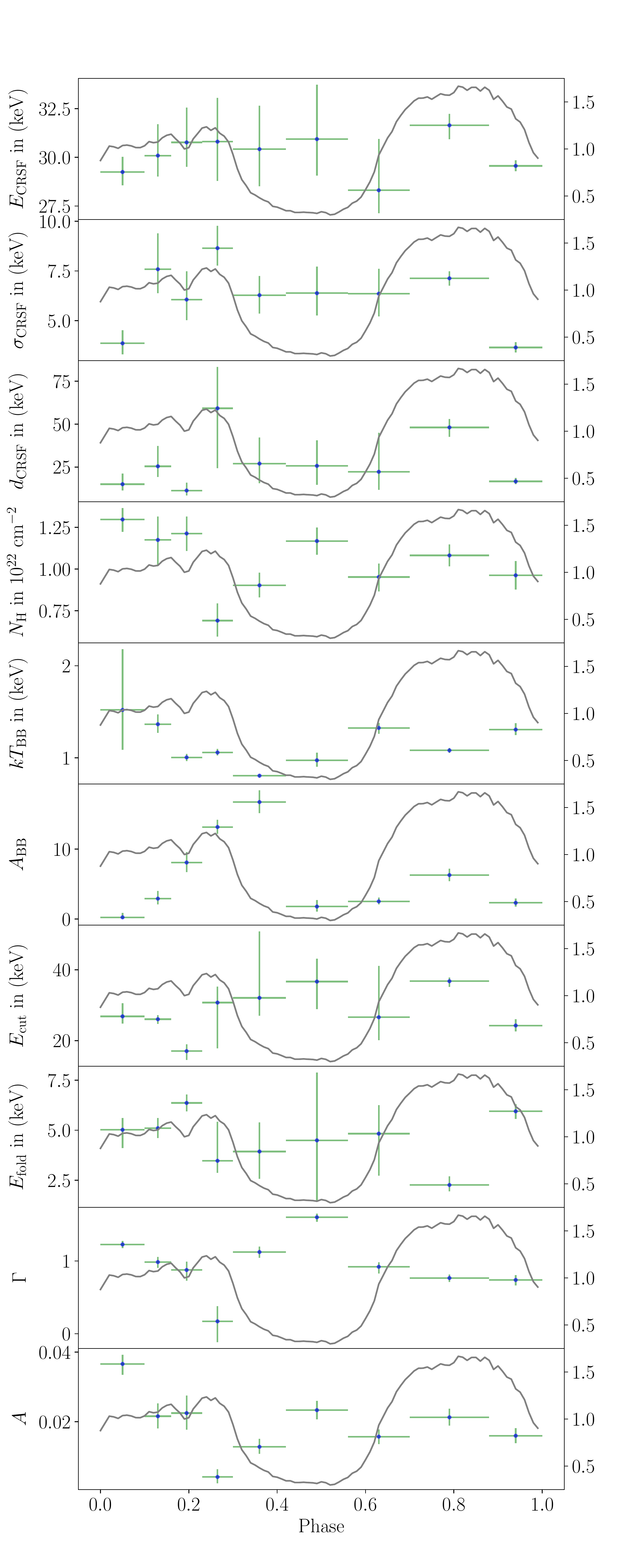}
\caption{Variation of parameters across pulse phase for \obs1 and \obs2. The left panels show the result for \obs1 and the right panels for \obs2. The pulse profile in 3-80~keV range for respective observation is shown as a gray line. The most notable feature is that the variation in the cyclotron energy with phase for \obs1\ matches the pulse profile. In \obs2 except for phases `e' and `f', the line energy follows the trend of the pulse profile as well. $N_{\rm{H}}$ and $\Gamma$ show correlated variability, which may be affected by intrinsic model degeneracy (see \autoref{fig:cont_plot}). The unit for $A_{\rm{BB}}$ is km$^2/(10~\rm{kpc})^2$ and for A is photons~cm$^{-2}$~s$^{-1}$~keV$^{-1}$ at 1~keV.}
\label{fig:phase_var}
\end{figure*}

The variation of the free parameters is tabulated in \autoref{tab:all_phases} and depicted in \autoref{fig:phase_var}. We see clear variations in model parameters with pulse phase, which can be understood as a effect of variation of magnetic field strength and other parameters along the different lines of sight probed for different phases. In \obs1, the variation of $E_\mathrm{CRSF}$ with the pulse phase follows the trend of the pulse profile. The Pearson R correlation coefficient for the line parameters with the corresponding value of the pulse profile are tabulated in \autoref{tab:pearson_coeff}. Cyclotron line parameters clearly vary with phase in both observations. The variation of cyclotron line central energy ($E_\mathrm{CRSF}$), width ($\sigma_\mathrm{CRSF}$), and depth ($d_\mathrm{CRSF}$) seems to be correlated with total flux in \obs1. However, no such correlation is seen in \obs2\ in the declining phase of the outburst. Excluding the data from phase `e' and `f' for \obs2 leads to a better correlation between the intensity at the pulse phase and cyclotron line parameters. 

The temperature of the blackbody for phase `a' seems to be smaller than the rest of the phase bins and the phase-averaged value. The area of the emitting region (derived from $A_{\rm{BB}}$) is also larger than in the other phase bins. To measure the effect of the temperature on the line parameters, we refit the phase bin spectrum while holding the temperature fixed at the phase-averaged value. The normalization of the component was left free. The line parameters thus obtained are consistent with the values reported in \autoref{tab:all_phases} and the normalization converges to the values suggested by the nearby phase bins. 

\begin{table}
\caption{The Pearson R correlation coefficient of the line parameters with the normalised pulse profile intensity for different phases. For \obs2 the value in parentheses indicate the coefficient computed excluding the phases `e' and `f'.}
\label{tab:pearson_coeff}
\centering
\begin{tabular}{ccc}
\hline
 & \textbf{\obs1} & \textbf{\obs2} \\
\hline
$E_{\rm CRSF}$ & 0.557 & 0.147 (0.783) \\
$\sigma_{\rm CRSF}$ & 0.322 & 0.026 (0.119)\\
$d_{\rm CRSF}$ & 0.408 & 0.230 (0.346)\\ \hline
\end{tabular}
\end{table}

\begin{figure*}\centering
\includegraphics[width =0.9\columnwidth]{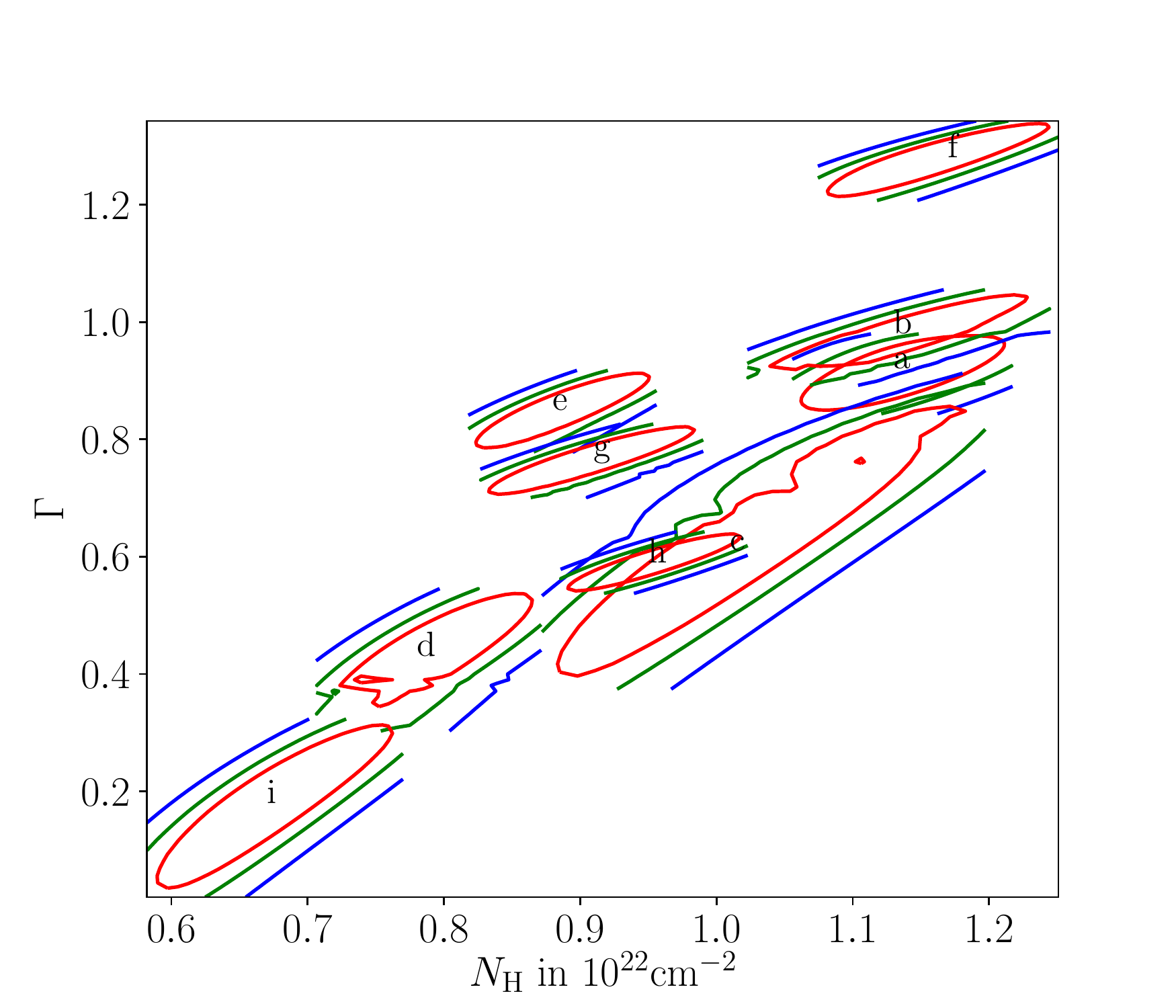} \includegraphics[width=0.9\columnwidth]{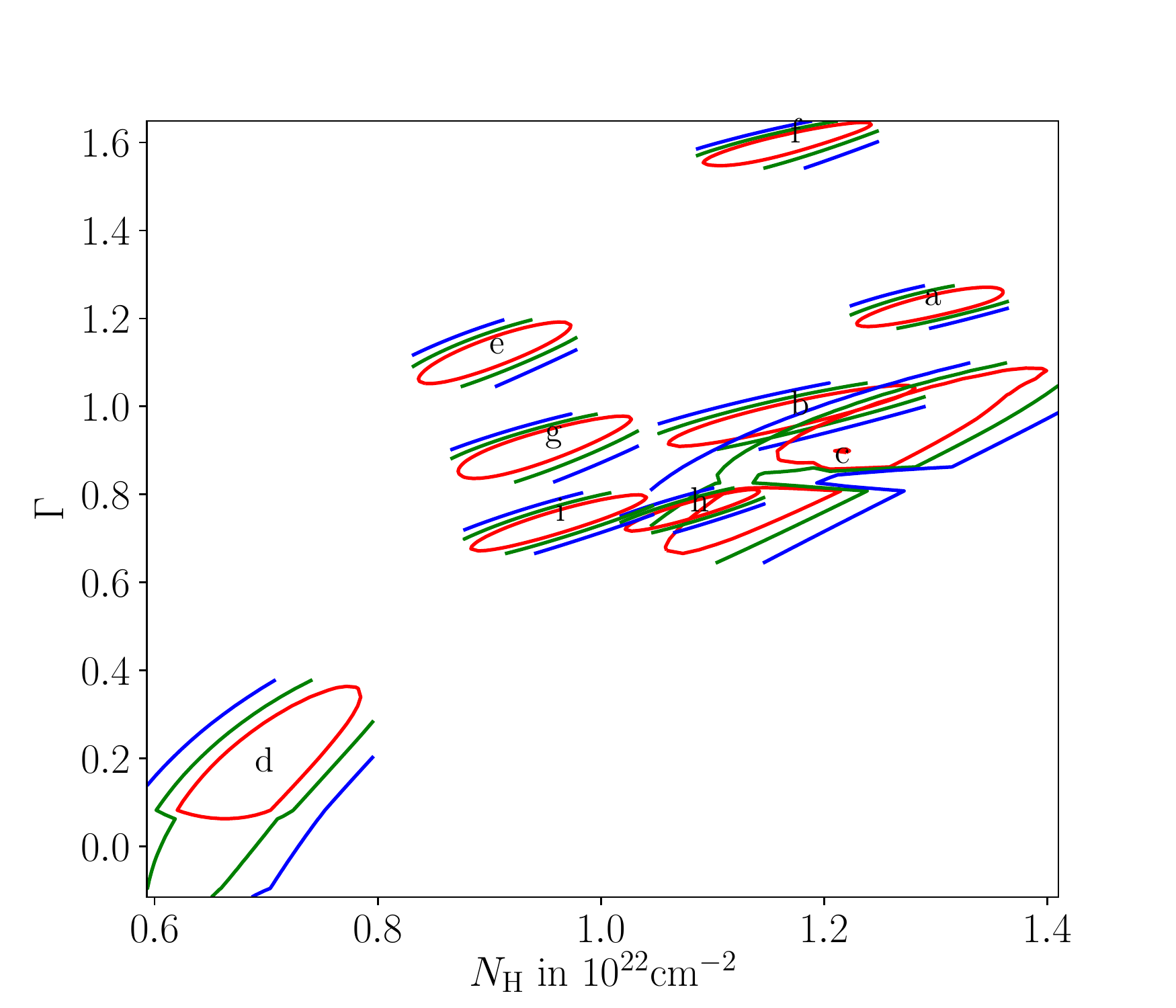}
\caption{The variation of $N_{\rm{H}}$ with $\Gamma$. The left panel shows the contour plot for \obs1 while the right panel shows the contour plot for \obs2. Each set of contours shows the degeneracy between these parameters for various pulse phases, indicated by a letter at the minimum $\chi^2$. Red, green and blue contours correspond to $\Delta \chi^2$ of 2.3, 4.6 and 9.2 respectively. The degeneracy in the best-fit values for $N_{\rm{H}}$ and $\Gamma$ is evident. However, the full range of variations across all phase bins exceeds the degenerate values of individual bins, and can therefore be interpreted as a real change in the source parameters. }
\label{fig:cont_plot}
\end{figure*}

We also find a dip in the absorbing column density ($N_\mathrm{H}$) at the peak in the pulse profile at phase 0.25, followed by a rise in $N_\mathrm{H}$ at the intensity dip at phase 0.5. This feature does not fully explain the energy-resolved lightcurves, which show large intensity variations even at high energies. 
We also note that the power-law index $\Gamma$ and normalisation $A$ show the same phase-resolved trend as $N_\mathrm{H}$. A correlated variation in the absorbing column and power-law parameters is often seen in fits. We explore this further by calculating $\chi^2$ contours for best-fit models obtained at different values of $N_\mathrm{H}$ and $\Gamma$ for each phase (see \autoref{fig:cont_plot}). The plot clearly shows the degeneracy in these two parameters for the fits at each phase. However, we also note that the total range of variations in $\Gamma$ and $N_\mathrm{H}$ is higher than the spread seen in individual phases and can therefore be interpreted as a real change in the source parameters.

Different continuum models have an effect on the line parameters. \cite{Muller2013A&A...551A...6M} showed that modelling the continuum shape using a Cutoff power-law~(\sw{CPL}) can result in sharper CRSF features compared to \sw{FDCUT+PL}. To test whether the wide CRSF parameters were an artifact of the continuum model used here, we also applied the \sw{CPL} and Negative and Positive power-law with Exponential cutoff \citep[\sw{NPEX},][]{Makishima1999ApJ...525..978M} to the phase resolved data. We find that the line parameters obtained in different continuum models are consistent with each other. Changing continuum model did also not warrant inclusion of the second absorption line in the spectral modelling. The reduction in $\chi^2$ is similar to the values obtained with \sw{FDCUT+PL}.  

\subsection{Adding the models}  \label{subsec:add_model}
\begin{figure}
\centering
\includegraphics[width=\columnwidth]{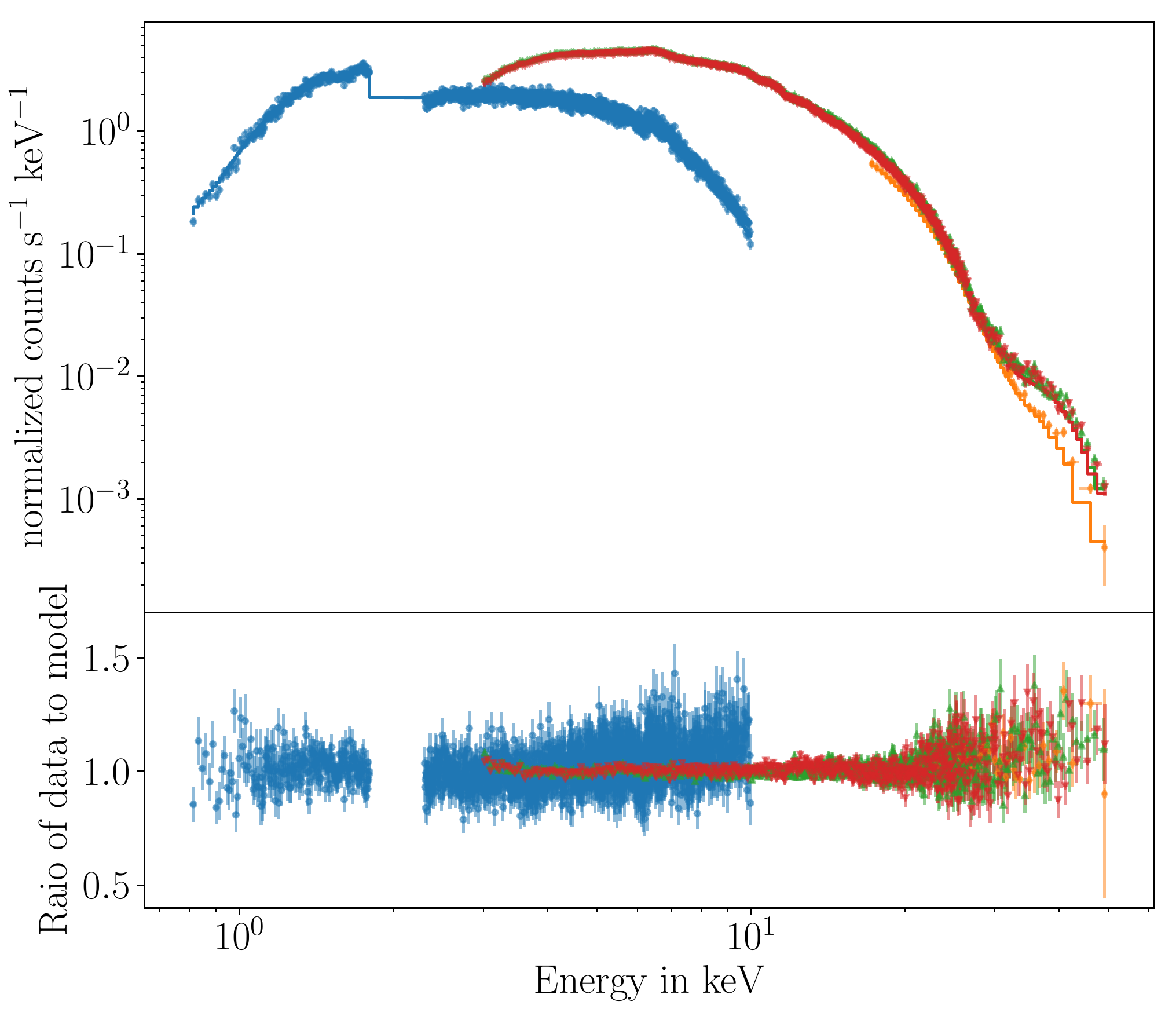}
\caption{The observed phase averaged spectrum for \obs1, plotted using the same colour scheme as  \autoref{fig:spec_all}. We compare this data with the ``combined model'' from \S\ref{subsec:add_model} without any refitting, and obtain a $\chi^2$ value of 4956.05 with 4199 degrees of freedom. This combined model is over-plotted on data in the top panel, while the bottom panel shows the ratio of the data to this model (without refitting).  Slight excess of the residuals can be seen in XIS from 7-10~keV which can be explained from the slight differences in the parameters between the respective best fits to the phase averaged data and the simulated spectrum from the addition of the models (\autoref{tab:ph_av_comp}).  }
\label{fig:comb_on_avg}
\end{figure}

The clear variation of cyclotron line parameters with phase (\autoref{fig:phase_var}) leads us to the question if the asymmetric line profile of the phase-averaged spectrum could simply arise by attempting to fit a single model to the phase-averaged spectrum. To test this hypothesis, we took our best-fit models for each phase (\S\ref{subsec:phaseres}), weighted them by the exposure of each phase bin, and added them together to create a single model. We then simulated a spectrum from this ``combined'' model using the \texttt{XSPEC} \texttt{fakeit} command. The simulated spectrum was fit with the complete two-component model used by \felix\ and in our phase-averaged analysis (\S\ref{subsec:phase_av}). We find that the spectral parameters obtained for this combined model spectrum are fully consistent with those obtained from the phase-averaged data (\autoref{tab:ph_av_comp}, ``Combined model'' columns). This striking agreement in the best-fit values indicates that the variation of cyclotron line parameters with phase is a sufficient explanation for the asymmetry in the phase-averaged line profile in \cep. For \obs1, the combined model when applied to the phase-averaged data without re-fitting results in a $\chi^2$ of 4956.05 with 4199 degrees of freedom. The phase averaged spectrum and the combined model have been shown in \autoref{fig:comb_on_avg}.

\begin{figure}
\centering
\includegraphics[width=1\columnwidth]{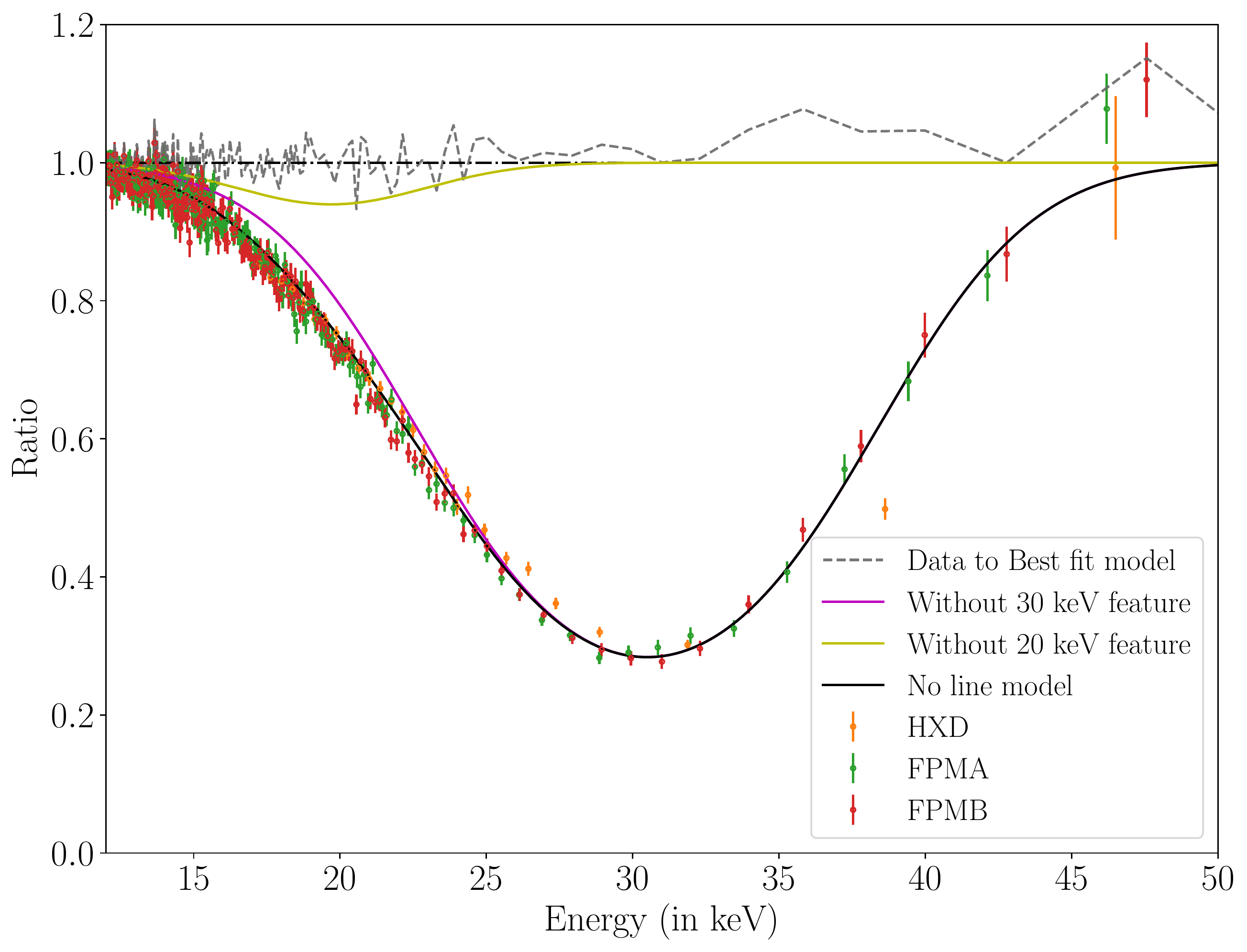}
\caption{
We create a synthetic spectrum from our ``combined model'' (\S\ref{subsec:add_model}) for \obs1, and fit it with the complete two-line model used in the phase-averaged analysis (\S\ref{subsec:phase_av}). The dashed gray line shows the ratio of synthetic data to the best-fit two-line model. The flatness of the line shows that the model is a good fit to synthesized data. Next, we set the depths of the two line components to zero, and plot ratios of data to this continuum-only model (without refitting). These data points follow the same colour scheme as \autoref{fig:spec_all}. A solid black line shows the ratio of this continuum-only model to the complete two-line model. Similarly, the yellow, and magenta lines correspond to models without the 20~keV feature and 30~keV features respectively.}
\label{fig:ratio}
\end{figure}

In this synthesized ``combined model'' spectrum, the cyclotron line is clearly asymmetric (\autoref{fig:ratio}). Notably the fitted parameters  in the combined model spectrum are close to the values reported in \felix.

This strongly suggests that the inherent asymmetry detected in the phase-averaged data is explained by the variation of spectral parameters with phase. We note that both single and double-line spectral models give much lower $\chi^2$ values when fit to our synthetic data derives from the combined model as compared to the actual phase-averaged data, so there may be small systematic effects in our combining procedure. The improvement obtained by switching from the single-line to double-line model for phase averaged data is much higher ($\Delta \chi^2 \sim 200$) than for the synthesised data ($\Delta \chi^2 \sim  40$), which could be attributed to two causes. First, it is possible that there is some inherent asymmetry in the line, which we cannot detect at high significance in the individual phase bins, but asserts itself in the phase-averaged data. The second possibility is that this smaller improvement is simply a result of the $\chi^2$ values for fits to synthesised data being already much lower than those for actual phase-averaged data. This interpretation is compatible with the excellent agreement in model parameters for synthesised and observed data. The current dataset will be unable to conclusively establish either of these cases.
 
\section{Discussion} \label{sec:disc}
We analysed \nus\ and \suz\ data of \cep\ obtained during the peak and decline of the 2014 outburst. The spectra are well fit by a model comprised of absorption, a blackbody component, a hard component, an iron line, and a cyclotron feature. We find that this data set gives identical results to those obtained from \nus\ and \emph{Swift}-XRT data~\pfelix. 
\cite{Jaisawal2015MNRAS.453L..21J} and \cite{Vybornov2017A&A...601A.126V} have reported a slightly anharmonic feature of the CRSF. Due to the spectrum being background dominated in this regime, we have excluded \nus\ data beyond 50~keV and \suz\ XIS data beyond 45~keV in our analysis.

We detect strong pulsations with an $\approx 66.33$~s period in both observations. The pulse profile varies with energy, and in general is smoother at high energies. The pulse fraction slightly increases with the energy as can be seen from the increase in both fractional amplitude and r.m.s. (see \autoref{fig:en_res_pulse}).

We undertake phase-resolved spectroscopy for both epochs. Data for individual phases are well-fit by a single cyclotron line, with no evidence for an asymmetric line profile like the one seen in phase-averaged spectra. We find clear evidence for variation of spectral parameters with phase. In particular, the cyclotron line energy and width show strong variations. These parameters are correlated with pulse intensity in \obs1 (at the peak of the outburst), but no obvious correlations are not seen in \obs2 (outburst decline phase). 

We simulate a phase-averaged spectrum by using the weighted sum of the spectral models for each phase. On attempting to fit this ``combined model''-based spectrum with our asymmetric line spectral model, we get an excellent fit. The derived parameters are fully consistent with those obtained by fitting the same model to the actual phase-averaged data. 
Modelling by \citet{Schwarm2017A&A...601A..99S} indicates that the variation in the parameters can be expected even if the emitting column is uniform. The fit by \cite{Schwarm2017A&A...601A..99S} to the asymmetric profile of \cep\ indicate that the asymmetry can also be explained by assuming a slight anharmonic spacing of the Landau levels coupled with the boosting of the photons due to relativistic electrons. In comparison to this, our work provides the evidence that the asymmetric profile can be obtained after averaging the variation of the parameters seen in different lines of sight. 
The variation of the physical conditions on the NS surface can also lead to variation in the spectral parameters across the phase bins \citep{Mukherjee2012MNRAS.420..720M}. 

While an asymmetric line provides a better fit to the combined model, the change in the $\chi^2$ value is much smaller than the one obtained for actual phase-averaged data. This effect can arise either due to the line having small, undetectable asymmetry in individual phase bins, or due the generally lower $\chi^2$ values obtained for synthesised data. Our current datasets are inadequate to rule out either hypothesis.

Such non-gaussian shapes for CRSFs are predicted by several models \citep[and references therein]{Mukherjee2012MNRAS.420..720M, Schwarm2017A&A...597A...3S, Schwarm2017A&A...601A..99S}, and \nus\ has the necessary sensitivity and energy resolution for obtaining data to test these models. However, comparison between models and data warrants good measurements of the line parameters and the underlying continuum. Our work shows that the cyclotron features and the continuum component are both strongly dependent on pulse phase in \cep, and any attempts to fit models to data should properly account for these variations. \cep\ is a promising start in this direction.

\section*{Acknowledgements}

This work has made use of data from the \nus\ mission, a project led by the California Institute of Technology, managed by the Jet Propulsion Laboratory, and funded by the National Aeronautics and Space Administration. We thank the \nus\ Operations, Software and Calibration teams for support with the execution and analysis of these observations.
This research has made use of the \nus\ Data Analysis
Software (\sw{NuSTARDAS}) jointly developed by the ASI Science Data Center (ASDC, Italy) and the California Institute of Technology (USA). R.B. acknowledges support by DLR grant 50 OR 1410.

\bibliography{ref.bib}

\begin{landscape}

\begin{table}
\centering
\caption {Comparing our phase averaged fit to \nus\ and \suz\ data with the \nus\ and \swift\ analysis results of \felix. The phase averaged spectrum is described in \S~\ref{subsec:phase_av} while the combined model spectrum is described in \S~\ref{subsec:add_model}. The parameters are listed for the model adopted by \felix\ which is used for fitting both phase-averaged spectrum and the combined model spectrum. The discrepancy in the blackbody norms ($A_{\rm BB}$) arises from our adoption of the \texttt{bbodyrad} model in \texttt{XSPEC}, as opposed to the \texttt{blackbody} function used by \felix. For completeness, we also list the $\chi^2$ values obtained for a model with a single absorption line. We caution that the parameter values in the table are for the two-line model, and different values are obtained for the single-line model.
}
\label{tab:ph_av_comp}

\begin{tabular}{c|cccc|ccc}
\hline 
Component  & Parameter  & \multicolumn{3}{|c|}{\obs1} & \multicolumn{3}{c}{\obs2 }\\
\hline 
 &  & \felix  & \multicolumn{2}{c|}{Current Work} & \felix  & \multicolumn{2}{c}{Current Work}\\
 
 &  & & Phase averaged  & Combined model & & Phase averaged  & Combined model \\
\hline 
Interstellar absorption  & $N_{\rm{H}}(10^{22} \rm{cm}^{-2})$  & $1.05_{-0.12}^{+0.11}$  & $1.06_{-0.04}^{+0.03}$ & $1.02\pm0.04$  & $1.41\pm 0.25$  & $1.09_{-0.03}^{+0.03}$ & $0.97\pm0.04$\\
\hline 
 & ${A_{\rm{cont}}}^a$  & $0.061_{-0.010}^{+0.008}$  & $0.066\pm0.004$ & $0.060\pm0.004 $  & $0.021_{-0.005}^{+0.004}$  & $0.0218\pm0.0007$ & $0.022\pm0.001$\\
Power law with  & $\Gamma$  & $0.83_{-0.11}^{+0.07}$  & $0.88_{-0.04}^{+0.03}$ & $ 0.80_{-0.06}^{+0.05}$  & $0.96_{-0.14}^{+0.09}$  & $1.00_{-0.03}^{+0.02}$ & $0.76_{-0.04}^{+0.03}$\\
Fermi Dirac cutoff & $E_{\rm{cut}}$(keV)  & $24\pm4$  & $26\pm2$ & $22_{-2}^{+3}$  & $25\pm 4$  & $28\pm2$ & $26_{-2}^{+3}$\\
 & $E_{\rm{fold}}$(keV)  & $5.7_{-0.6}^{+0.5}$  & $5.3_{-0.5}^{+0.4}$ & $5.9_{-0.5}^{+0.3}$  & $5.7_{-0.8}^{+0.6}$  & $4.9_{-0.6}^{+0.5}$ & $4.9_{-0.6}^{+0.5}$\\
\hline 
Cyclotron absorption  & $E_{\rm{CRSF}}$(keV)  & $30.39_{-0.14}^{+0.17}$  & $30.52_{-0.14}^{+0.18}$ & $ 30.5_{-0.14}^{+0.22}$  & $29.42_{-0.24}^{+0.27}$  & $29.5_{-0.2}^{+0.3}$ & $29.5_{-0.2}^{+0.3}$\\
feature 1 & $\sigma_{\rm{CRSF}}$(keV)  & $5.8\pm0.4$  & $6.1\pm0.3$ & $ 5.7_{-0.3}^{+0.4}$ & $4.9\pm0.4$  & $5.3\pm0.3$ & $6.1\pm0.4$ \\
 & $d_{\rm{CRSF}}$(keV)  & $20_{-4}^{+5}$  & $24_{-3}^{+4}$ & $ 18_{-2}^{+4}$  & $16.6_{-3.0}^{+4.0}$  & $21_{-3}^{+4}$ & $23_{-4}^{+6}$ \\
\hline 
Cyclotron absorption  & $E_{\rm{abs}}$(keV)  & $19.0_{-0.4}^{+0.5}$  & $19.0_{-0.3}^{+0.4}$ & $ 19.7_{-0.9}^{+1.4}$  & $18.5\pm0.7$  & $18.0_{-0.5}^{+0.6}$ & $17.8_{-0.6}^{+0.8}$ \\
feature 2 & $\sigma_{\rm{abs}}$(keV)  & $2.5\pm0.4$  & $2.6\pm0.3$ & $ 3.2_{-0.9}^{+1.1}$  & $2.1\pm0.5$  & $2.0\pm0.4$ & $2.5_{-0.6}^{+0.7}$ \\
 & $d_{\rm{abs}}$(keV)  & $0.60_{-0.17}^{+0.24}$  & $0.68_{-0.17}^{+0.24}$ & $ 0.5_{-0.3}^{+0.9}$ & $0.37_{-0.15}^{+0.21}$  & $0.36_{-0.12}^{+0.16}$ & $0.4_{-0.2}^{+0.3}$ \\
\hline 
Iron line emission & $A$(Fe K$\alpha$)$^a$  & $\left(1.39_{-0.14}^{+0.16}\right)\times10^{-3}$  & $\left(1.11_{-0.09}^{+0.10}\right)\times10^{-3}$ & $\left(1.10_{-0.07}^{+0.08} \right)\times10^{-3}$ & $\left(2.8_{-0.6}^{+0.8}\right)\times10^{-4}$  & $\left(1.8\pm0.3\right)\times10^{-4}$ & $\left(1.6\pm0.4\right)\times10^{-4}$ \\
 & $\sigma$(Fe K$\alpha$)(keV)  & $0.42\pm0.05$  & $0.32_{-0.03}^{+0.04}$ & $0.29\pm0.02$ & $0.34_{-0.10}^{+0.12}$  & $0.19\pm0.07$ & $0.17_{-0.10}^{+0.08}$ \\
 & $E$(Fe K$\alpha$)(keV)  & $6.47\pm0.03$  & $6.49\pm0.02$ & $6.48\pm0.02 $ & $6.39_{-0.07}^{+0.06}$  & $6.42\pm0.03$ & $6.43\pm0.05$ \\
\hline 
Blackbody & $A_{\rm{BB}}$  & $\left(2.22_{-0.29}^{+0.41}\right)\times10^{-3}$  & $19\pm2^{b}$ & $25\pm3^{b}$& $\left(7.3_{-1.3}^{+1.7}\right)\times10^{-4}$  & $3.8\pm0.4^{b}$ & $3.8_{-0.6}^{+0.7~b}$ \\
 & $kT_{\rm{BB}}$(keV)  & $0.90\pm0.03$  & $0.960\pm0.017$ & $0.92\pm0.02$ & $0.96\pm0.06$  & $1.09\pm0.02$ & $1.04_{-0.04}^{+0.03}$ \\
\hline 
Detector  & $C_{\rm{FPMB}}$  & $1.0319\pm0.0019$  & $1.029\pm0.001$  & $1.029\pm0.001$ & $1.023\pm0.004$  & $1.019\pm0.002$ & $1.016\pm0.002$  \\
normalization  & $C_{\rm{XRT/XIS}}$ & $0.962\pm0.019$  & $0.891\pm0.002$  & $0.883\pm0.002$ & $0.91\pm0.05$  & $0.900\pm-0.002$ & $0.884\pm0.002$ \\

& $C_{\rm{HXD}}$ &  & $1.246\pm0.004$  & $1.141\pm 0.002 $ &   & $1.249\pm0.004$ & $1.187\pm0.005$ \\
\hline 
$\chi^{2}$/dof  & Two line model & 1215.83/1082  & 4708.28/4199 & 4045.28/4192 & 776.35/670 & 4073.38/3794 & 3968.90/3926 \\
$\chi_{\rm{red}}^{2}$ &  & 1.124  & 1.121 & 0.965 & 1.159 & 1.073 & 1.011\\ \hline
$\chi^{2}$/dof  & One line model & 1324/1085 & 4903.03/4202 & 4085.72/4195 & 802/673 & 4127.85/3797 & 4015.21/3929 \\
$\chi_{\rm{red}}^{2}$ &  & 1.22  & 1.166 & 0.973 & 1.19 & 1.087 & 1.022\\
\hline

\end{tabular}
\flushleft
$^a$ in photons keV$^{-1}$ s$^{-1}$ cm$^{-2}$ at 1~keV

$^b$ in km$^2/(10~\rm{kpc})^2$ 
\end{table}
\end{landscape}

\begin{landscape}
\begin{table}
\centering

\caption{Spectral parameters from phase-resolved analysis of \cep\ fitted with the model \texttt{phabs*gabs*(bbodyrad+gaussian+fdcut*powerlaw)}. The variation of these parameters is depicted in Figure~\ref{fig:phase_var} and the results are discussed in \S~\ref{subsec:phaseres}}
\label{tab:all_phases}

\scriptsize

\begin{tabular}{|l|c|c|c|c|c|c|c|c|c|c|}
\hline 
\multicolumn{2}{|c|}{ Phase bin} 				& a 						& b 						& c 						& d 						& e 						& f 						& g 						& h 						& i \\
\hline 		
\multicolumn{2}{|c|}{Start phase} & 0 & 0.1 & 0.16 & 0.23 & 0.30 & 0.42 & 0.56 & 0.70 & 0.88\\ \hline
\multicolumn{2}{|c|}{End phase} & 0.1 & 0.16 & 0.23 & 0.30 & 0.42 & 0.56 & 0.70 & 0.88 & 1\\
\hline
\hline
\multicolumn{2}{|c|}{ Parameters} & \multicolumn{9}{|c|}{ Observation 1} \\
\hline 		
Interstellar absorption & $N_{\rm{H}}$ ($10^{22}$~cm$^{-2}$)  		&	$1.13\pm0.08$ & $1.13\pm0.11$ & $1.01_{-0.14} ^{+0.18}$ & $0.78\pm0.08$ & $0.88\pm0.07$ & $1.17_{-0.09} ^{+0.08}$ & $0.91\pm0.08$ & $0.95\pm0.07$ & $0.67\pm0.09$ \\
\hline 		
Blackbody & ${A_{\rm{BB}}}^{a}$     & $75 _{-15} ^{+18}$ & $12\pm5$ & $49_{-14} ^{+10}$ & $63\pm4$ & $52\pm4$ & $15\pm3$ & $10.7_{-1.7} ^{+1.8}$ & $12.2\pm1.6$ & $23\pm3$ \\
%\hline 		
 & $kT_{\rm{BB}}$(keV) 			& $0.66\pm0.04$ & $0.99\pm0.08$ & $0.91_{-0.03} ^{+0.05}$ & $0.97\pm0.02$ & $0.90\pm0.03$ & $0.95_{-0.04} ^{+0.03}$ & $1.08_{-0.04} ^{+0.03}$ & $1.13\pm0.03$ & $1.05\pm0.03$ \\
\hline 		
Power law & ${A_{\rm{cont}}}^{b}$  			& $0.069_{-0.008} ^{+0.007}$ & $0.099_{-0.015} ^{+0.016}$ & $0.064_{-0.016} ^{+0.022}$ & $0.034_{-0.007} ^{+0.009}$ & $0.050\pm0.007$ & $0.071\pm0.009$ & $0.041\pm0.006$ & $0.045\pm0.005$ & $0.027_{-0.006} ^{+0.007}$ \\
%\hline 			
with Fermi Dirac & $\Gamma$  				& $0.92_{-0.08} ^{+0.06}$ & $0.98\pm0.08$ & $0.61_{-0.24} ^{+0.30}$ & $0.43_{-0.13} ^{+0.11}$ & $0.85_{-0.07} ^{+0.06}$ & $1.28 _{-0.08} ^{+0.06}$ & $0.77_{-0.07} ^{+0.06}$ & $0.59\pm0.05$ & $0.18_{-0.16} ^{+0.14}$ \\
%\hline 			
cut off & $E_{\rm{cut}}$(keV) 			& $27 _{-4} ^{+11}$ & $38 _{-12} ^{+3}$ & $14 _{-4} ^{+17}$ & $30_{-13} ^{+6}$ & $31 _{-7} ^{+9}$ & $29 _{-4} ^{+7}$ & $33 _{-7} ^{+4}$ & $38.6 _{-0.9} ^{+0.7}$ & $12\pm2$ \\
%\hline 			
 & $E_{\rm{fold}}$(keV)			& $5.5_{-2.6} ^{+0.6}$ & $2.9_{-0.7} ^{+3.6}$ & $6.4 _{-3.8} ^{+0.4}$ & $3.9 _{-0.8} ^{+1.9}$ & $4.2 _{-1.7} ^{+1.5}$ & $4.6 _{-1.6} ^{+0.9}$ & $4.1 _{-0.9} ^{+1.6}$ & $2.33_{-0.19} ^{+0.22}$ & $6.86\pm0.17$ \\
\hline 			
Cyclotron & $E_{\rm{CRSF}}$(keV)			& $31.7_{-0.4} ^{+1.7}$ & $34_{-4} ^{+1.5}$ & $30.8_{-0.9} ^{+1.1}$ & $31.5_{-1.3} ^{+2.3}$ & $31.5_{-1.1} ^{+4.9}$ & $28.7_{-1.0} ^{+2.1}$ & $31.8_{-1.1} ^{+1.7}$ & $33.4_{-0.5} ^{+0.4}$ & $30.43\pm0.17$ \\
%\hline 	
Absorption & $\sigma_{\rm{CRSF}}$(keV)  	& $6.3 _{-0.5} ^{+0.9}$ & $9.2_{-2.8} ^{+0.9}$ & $6.9_{-1.1} ^{+1.9}$ & $9.1_{-0.5} ^{+1.2}$ & $8.0_{-0.5} ^{+1.3}$ & $5.9_{-0.9} ^{+1.0}$ & $8.1_{-0.5} ^{+0.7}$ & $8.4\pm0.3$ & $4.02_{-0.16} ^{+0.18}$ \\
%\hline 		
line & $d_{\rm{CRSF}}$(keV) 		& $28 _{-6} ^{+24}$ & $68_{-41} ^{+15}$ & $16_{-5} ^{+31}$ & $61_{-31} ^{+29}$ & $42_{-16} ^{+32}$ & $15_{-6} ^{+16}$ & $46_{-16} ^{+15}$ & $65\pm4$ & $14.4_{-0.8} ^{+0.9}$ \\
\hline 		
																								
	&	$\chi^2/$d.o.f. & $2470.23/2372$ & $2082.82/2010$ & $2349.73/2269$ & $2431.45/2373$ & $2506.83/2580$ & $2125.74/2167$ & $2635.04/2538$ & $3393.67/3284$ & $2854.08/2786$ \\
\hline 
\hline
\multicolumn{2}{|c|}{ Parameters} & \multicolumn{9}{|c|}{ Observation 2} \\
\hline 		
Interstellar absorption & $N_{\rm{H}}$ ($10^{22}$~cm$^{-2}$)  				& $1.29\pm0.07$ & $1.17_{-0.15} ^{+0.14}$ & $1.21\pm0.10$ & $0.69_{-0.09} ^{+0.10}$ & $0.90_{-0.07} ^{+0.08}$ & $1.17\pm0.08$ & $0.95_{-0.09} ^{+0.08}$ & $1.08\pm0.06$ & $0.96\pm0.08$ \\

\hline 		
Blackbody & ${A_{\rm{BB}}}^{a}$ 		& $0.24_{-0.19} ^{+0.62}$ & $2.9_{-0.8} ^{+1.1}$ & $8.1_{-1.3} ^{+1.5}$ & $13.1_{-0.9} ^{+1.0}$ & $16.7\pm1.6$ & $1.8_{-0.7} ^{+0.9}$ & $2.5_{-0.4} ^{+0.5}$ & $6.3\pm0.9$ & $2.3_{-0.5} ^{+0.6}$ \\
%\hline 		
 & $kT_{\rm{BB}}$(keV) 			& $1.5_{-0.4} ^{+0.6}$ & $1.36_{-0.09} ^{+0.10}$ & $1.00\pm0.04$ & $1.06_{-0.03} ^{+0.04}$ & $0.80\pm0.02$ & $0.97_{-0.07} ^{+0.08}$ & $1.32_{-0.06} ^{+0.05}$ & $1.08\pm0.03$ & $1.30_{-0.06} ^{+0.07}$ \\
\hline 		
Power law & ${A_{\rm{cont}}}^{b}$  			&  $0.036_{-0.003} ^{+0.002}$ & $0.021_{-0.003} ^{+0.004}$ & $0.022\pm0.005$ & $0.0041 _{-0.0018} ^{+0.0022}$ & $0.013\pm0.002$ & $0.023\pm0.003$ & $0.015\pm0.002$ & $0.021\pm0.002$ & $0.016\pm0.002$ \\
%\hline 			
with Fermi Dirac & $\Gamma$  			&	$1.23\pm0.05$ & $0.98_{-0.08} ^{+0.07}$ & $0.87_{-0.15} ^{+0.12}$ & $0.17_{-0.28} ^{+0.21}$ & $1.12\pm0.07$ & $1.60_{-0.06} ^{+0.05}$ & $0.92_{-0.09} ^{+0.06}$ & $0.76\pm0.05$ & $0.74_{-0.07} ^{+0.06}$ \\
%\hline 			
cut off & $E_{\rm{cut}}$(keV) 			&  $27 _{-2} ^{+4}$ & $26.0_{-1.3} ^{+1.2}$ & $17.0_{-2.4} ^{+1.9}$ & $30_{-13} ^{+4}$ & $32_{-5} ^{+19}$ & $36_{-8} ^{+6}$ & $27_{-6} ^{+14}$ & $36.7_{-1.6} ^{+1.1}$ & $24.2_{-1.6} ^{+1.8}$ \\
%\hline 			
 & $E_{\rm{fold}}$(keV)			 & $5.0_{-0.9} ^{+0.6}$ & $5.1\pm0.5$ & $6.4\pm0.4$ & $3.5_{-0.6} ^{+1.9}$ & $3.9\pm1.4$ & $4.5_{-3.0} ^{+3.4}$ & $4.8_{-2.1} ^{+1.4}$ & $2.3_{-0.3} ^{+0.4}$ & $5.9\pm0.4$ \\
\hline 			
Cyclotron & $E_{\rm{CRSF}}$(keV)			& $29.2 _{-0.7} ^{+0.8}$ & $30.1 _{-1.1} ^{+1.6}$ & $30.8_{-1.2} ^{+1.8}$ & $30.8_{-2.0} ^{+2.2}$ & $30.4_{-1.9} ^{+2.2}$ & $30.9_{-1.9} ^{+2.8}$ & $28.3_{-1.2} ^{+2.6}$ & $31.6_{-0.7} ^{+0.6}$ & $29.6\pm0.3$ \\
%\hline 	
Absorption & $\sigma_{\rm{CRSF}}$(keV)  & $3.9_{-0.5} ^{+0.6}$ & $7.5_{-1.2} ^{+1.8}$ & $6.0_{-1.0} ^{+1.4}$ & $8.6_{-0.8} ^{+1.1}$ & $6.3\pm0.9$ & $6.3_{-1.1} ^{+1.3}$ & $6.3_{-1.1} ^{+1.2}$ & $7.1\pm0.4$ & $3.6_{-0.2} ^{+0.3}$ \\
%\hline 		
line & $d_{\rm{CRSF}}$(keV) 		& $15_{-3} ^{+6}$ & $25 _{-6} ^{+11}$ & $11_{-3} ^{+4}$ & $59_{-34} ^{+24}$ & $27_{-11} ^{+15}$ & $25_{-11} ^{+14}$ & $22_{-10} ^{+22}$ & $48\pm5$ & $16.6_{-1.7} ^{+1.9}$ \\
\hline 		
 & $\chi^2/$d.o.f & $2255.14/2285$ & $1736.34/1646$ & $1898.66/1962$ & $1930.98/2003$ & $1934.30/1930$ & $1695.34/1646$ & $2495.15/2546$ & $3166.90/3159$ & $2743.86/2716$ \\												
 
\hline 
\end{tabular}
\flushleft

$^a$ in km$^2/(10~\rm{kpc})^2$ 

$^b$ in photons keV$^{-1}$ s$^{-1}$ cm$^{-2}$ at 1~keV

\end{table}
\end{landscape}

\end{document}